\documentclass[a4paper,11pt]{article}
\usepackage{jcappub} 

\arxivnumber{2411.06014}
\title{\boldmath
Non-linear Cosmological Perturbations for Coupled Dark Energy}
\author{Bilal T\"udes}

\author{and Luca Amendola}%

\affiliation{%
 Institut f\"ur Theoretische Physik, Universit\"at Heidelberg,\\
Philosophenweg 16, 69120 Heidelberg, Germany
}%
\emailAdd{tuedes@thphys.uni-heidelberg.de}
\emailAdd{l.amendola@thphys.uni-heidelberg.de}

\date{\today}

\abstract{
We derive the one-loop perturbation kernels for a minimal modified gravity model in which dark energy is coupled to dark matter via a constant coupling. We derive the time-dependent kernels via analytical and numerical solutions and provide accurate fitting functions. These kernels can be directly employed to test for modified gravity in forthcoming large-scale surveys.}

\usepackage{graphicx}
\usepackage{comment}

\usepackage{amsmath}
\usepackage{dcolumn}
\usepackage{bm}
\usepackage{xcolor}
\usepackage{hyperref}
\renewcommand{\eqref}[1]{(\ref{#1})}

\usepackage{subcaption}
\usepackage[labelfont=bf, font=small, width=0.8\textwidth]{subcaption}

\begin{document}

\maketitle
\flushbottom

\section{\label{sec:level1}Introduction}
Testing gravity at cosmological scales \cite{Amendola_2020,Martinelli_2021,cosmological_test_gr} is finally becoming possible thanks to several large-scale surveys like BOSS \cite{Dawson_2012}, DES \cite{DES}, DESI \cite{desi1}, Euclid \cite{euclid} and, in the near future \cite{Sprenger_2019}, LSST \cite{Zhan_2018} and SKA \cite{maartens2015,2020}. 

A common characteristic of most alternatives to Einstein's gravity is the introduction of a non-minimal coupling of a scalar field to matter or to the metric, thereby adding at least one new parameter to the cosmological set. In this paper we focus on what could be classified as the  simplest model of modified gravity: a constant coupling between a scalar field and dark matter. The scalar field can drive the cosmic acceleration, and therefore be a form of  dark energy,  
but this is not a necessary condition. This dark-dark coupling bypasses the stringent conditions of local gravity since ordinary matter (i.e. baryons) is left uncoupled and feels only standard gravity. We refer to this model simply as coupled dark energy (CDE) \cite{Amendola2000,Tocchini_Valentini_2002,Gumjudpai_2005,B_hmer_2008,Singh_2016,Bernardi_2017}.

Previous work  on the observable effects of this minimal model has focused on the linear regime (e.g. \cite{G_mez_Valent_2020,van_de_Bruck_2017,Yang_2016,Fay_2016,Xia_2009,Honorez_2010}), corresponding to very large scales, typically larger than 50 Mpc$/h$. It is however now clear that the mildly non-linear regime, reaching down to roughly 10 Mpc$/h$, contains a wealth of additional information, allowing, in particular, to break some degeneracies among cosmological parameters.

The  extension of cosmological data analysis to the non-linear regime can be performed in two ways: either resorting to $N$-body simulations, or by going  higher in perturbation theory. The first avenue is of course more powerful and is clearly preferable if one deals with a restricted class of models that can be reliably and efficiently simulated. In the present context of uncertainty about what could be the preferred class of models beyond $\Lambda$CDM, however, the perturbation approach still has some appeal, since it can  straightforwardly encompass large classes of models.

In this paper we consider therefore the next-to-leading order perturbation for a minimal dark-dark model, characterized by a coupling constant $\beta$ and a generic scalar field potential.
We consider two potentials: an exponential potential, as suggested by theoretical considerations \cite{Wetterich_2008}; and a linear approximation that is suitable when the scalar field does not change much during the relevant period (i.e., within the range of observations). The linear potential  has the advantage that an analytical solution for the background can be obtained.
In  the limit in which the potential reduces to a cosmological constant, this coupled model introduces a single constant beyond standard  and represents therefore a  minimal modified gravity model.

In higher-order perturbation theory, the perturbation variables are expressed as convolutions of the first order variables. The main difficulty consists  therefore in deriving the kernels of the convolutions. In this paper
we obtain the kernels for a generic cosmology following the approach of \cite{D_Amico_2021} and then we specialize our results to the CDE model, providing analytical and/or numerical forms of the kernels as a function of time, of the coupling parameter, and of the potential slope. These forms are ready to be used in future work to forecast the performance of cosmological surveys and to analyse real data. The analytical expressions are both useful to identify the general properties of the solutions and also very convenient for extensive MonteCarlo sampling of posteriors in real data analysis.

The structure of this paper is as follows. In Section~\ref{sec:level2}, we introduce the background equations for the coupled dark energy (CDE) model, rewrite them using the dimensionless parameters we define, and present the evolution equation for the linear growth-rate function $f$. In Section~\ref{sec:level3}, we present the conservation equations in the Newtonian limit and Fourier space, describe the perturbative approach to find the second- and third-order kernel functions, and then derive the evolution equations for the time-dependent kernel coefficients  using the continuity and Euler equations. In Section~\ref{sec:level4}, we solve these time-dependent coefficients for the CDE model. We begin by recalling the known solutions in the Einstein--de Sitter (EdS) case and then find the coefficients for the $\phi$-matter-dominated epoch. Finally,  assuming the scalar field evolves with a linear potential, we obtain analytical solutions for the background  parameters . In Section~\ref{sec:level6}, we summarize our findings and conclude. Appendix~\ref{Appendix:Fitting functions} provides fitting functions for the kernel coefficients in the CDE model with linear and exponential potentials, along with a discussion of their initial conditions for numerical integration.
\section{\label{sec:level2}Background and Perturbations}
In our model, the dark energy scalar field (subscript $\phi$) is coupled to the non-relativistic matter (subscript $m$) field in a flat FLRW background. Therefore, the individual energy-momentum tensors are not conserved due to interaction terms on the right-hand side:

\begin{equation}
\nabla_\mu T_{\nu (\phi)}^\mu=\beta T_m \nabla_\nu \phi \, , \quad 
\quad \nabla_\mu T_{\nu(m)}^\mu=-\beta T_m \nabla_\nu \phi \, .\label{eq:energy-mom}
\end{equation}
These equations are covariant and therefore can be applied to any metric, so no assumptions about the perturbation of the scalar field are necessary at this stage.
Here $\beta$ is a dimensionless constant coupling term that quantifies the strength of the dark-dark interaction, while  $T_m$ refers to the trace of the energy-momentum tensor for pressureless matter. 
Baryonic matter is assumed to be uncoupled so that the local gravity constraints are automatically satisfied.
In CDE, the gravitational equations remain the same as in standard gravity.
Other formulations of such interaction terms can be found in the literature \cite{Billyard_2000,Chen_2009,Lopez_Honorez_2010,Shahalam_2015,zhang}.

The coupling constant $\beta$ has been constrained, mainly by CMB, to be smaller than 0.04 roughly \cite{G_mez_Valent_2020}. However, this upper limit assumes a constant value across the entire cosmic evolution. A variable coupling might in general take larger values. Here, therefore, we will explore also values up to $\beta=0.2$.

The time component ($\nu=0$) of Eq. \eqref{eq:energy-mom} provides evolution equations for dark matter, dark energy and radiation (subscript $r$) respectively: 
\begin{align}
    \rho_{m}' +3(P_{m}+ \rho_{m})&=-\beta\rho_{m}\phi'
    \label{eq:matterdens},
\\  
 \rho_{\phi}' +3(P_{\phi}+ \rho_{\phi})&=\beta\rho_{m}\phi'
\label{eq:scalar_field_dens},
\\
    \rho_{r}' +4 \rho_{r}&=0 \, .
    \label{eq:rad}
\end{align}
The prime symbol indicates differentiation with respect to \(\eta = \log(a)\), where \(a\) denotes the scale factor. 
 The radiation Eq. \eqref{eq:rad} is independent of the coupling since its energy-momentum tensor is traceless.

The  Friedmann equation reads:
\begin{equation}
    3H^{2}= \rho_{\phi}+\rho_{m}+\rho_{r}\, .
    \label{eq:H/H'}
\end{equation}
The conservation equation for $\phi$ can also be written in the Klein-Gordon form:
\begin{equation}
\label{eq:KG}
    \phi''+(3+\frac{H'}{H})\phi'+\frac{1}{H^2}\frac{dV}{d\phi}=3\beta \Omega_c \, .
\end{equation}

We now rewrite the background equations in terms of dimensionless parameters. These parameters can be defined as follows:
\begin{equation}
x=\frac{\phi'}{\sqrt{6}},\quad y=\frac{1}{H}\sqrt{\frac{V}{3}},\quad z=\sqrt{\Omega_{r}},\quad v=\sqrt{\Omega_{b}} \, .
\label{eq:x,y,z,v}
\end{equation}

The quantities $\Omega_K= x^2 \), \( \Omega_P = y^2 \), \( \Omega_r = z^2 \), and \( \Omega_b = v^2 \) represent the fractions of total energy associated with field kinetic energy, field potential energy, radiation, and baryons, respectively. Additionally, the field energy fraction is given by \( \Omega_\phi = x^2 + y^2 \). The effective state parameter, defined as the total pressure-to-density ratio,  is given in terms of the parameters in \eqref{eq:x,y,z,v} as $w_{\rm eff} = x^2 - y^2 + \frac{z^2}{3}$;  while the field state parameter, \( w_\phi = \frac{P_{\phi}}{\rho_{\phi}} \), is expressed as \( w_\phi = \frac{x^2 - y^2}{x^2 + y^2} \).   We take the field potential to have an exponential form, \( V(\phi) = V_0 e^{-\mu \phi} \). Since the total   energy fractions must sum to one, the cold dark  matter energy density fraction is given by \( \Omega_c = 1 - x^2 - y^2 - z^2 -v^2\). 
Using the above relation, and rewriting our background equations in terms of the dimensionless parameters defined in Eq. (\ref{eq:x,y,z,v}), we obtain an autonomous dynamical system:
\begin{eqnarray}
x' && =\sqrt{\frac{3}{2}}\beta\left(1-v^{2}-x^{2}-y^{2}-z^{2}\right) +  \sqrt{\frac{3}{2}}\mu y^{2}+\frac{1}{2}x\left(3x^{2}-3y^{2}+z^{2}-3\right) \, ,
\label{eq:x}\\
y' && =-\sqrt{\frac{3}{2}}\mu xy-\frac{1}{2}y\left(-3x^{2}+3y^{2}-z^{2}-3\right)  \, ,
\label{eq:y}\\
v' && =-\frac{1}{2}v\left(-3x^{2}+3y^{2}-z^{2}\right)  \, ,
\label{eq:v}\\
z' && =-\frac{1}{2}z\left(-3x^{2}+3y^{2}-z^{2}+1\right)  \, ,
\label{eq:z}\\
H' &&=-\frac{1}{2}H\left(3x^{2}-3y^{2}+z^{2}+3\right)  \, , \label{eq:h}  
\end{eqnarray}
 where \(\mu =  -\frac{1}{V}\frac{\partial V}{ \partial \phi}\) is the potential slope parameter.
To gain insight into the dynamics of the system, one can identify the critical points of the phase space  and evaluate their stability conditions. This has been already investigated in detail and will not be repeated here
(see e.g. the review \cite{Bahamonde_2018}). However, for the scope of our work, we will specifically address one critical point, namely the existence of a matter-dominated era in which the scalar field is not negligible. This critical point, known as ${\phi}$MDE, will be detailed  in  Section \ref{sec:level42}.

\subsection{\label{sec:level22}Linear growth rate}

In this work we assume that baryonic growth is driven by the dark matter growth \cite{Amendola_2004}, implying $\delta_b'/\delta_b = \delta'_c/\delta_c\equiv f$.
From now on we neglect radiation since we study only the evolution of late-time cosmology.
At the linear level, we obtain then the following equation  for the growth rate $f$:
\begin{eqnarray}
    &&f'+  f^ {2}  +  \frac {1}{2}  (1-  3w_ {\rm eff} -2\beta \phi  ')f  -\frac {3}{2} [(1+2\beta^2)\Omega_c +\Omega_b] =0  \, ,\nonumber
\end{eqnarray}
which can be rewritten as
\begin{eqnarray}
    &&f'+  f^ {2}  +  Ff- S =0 \, ,
    \label{eq:f}
\end{eqnarray}
where, in terms of the dimensionless variables of Eq. \eqref{eq:x,y,z,v}, we have
\begin{eqnarray}
    F&=&\frac{1}{2}-\frac{3}{2}(x^2-y^2)-\sqrt{6}\beta x  \, , \label{eq:F}\\
     S&=&\frac{3}{2}[(1+2\beta^2)(1-x^2-y^2-v^2)+v^2]  \, .
    \label{eq:S}
\end{eqnarray}
Analytical solutions exist when  $\Omega_c,\Omega_b,\phi',w_{\rm eff}$ are constant.  Otherwise,  numerical computation of Eq. \eqref{eq:f} becomes necessary. A simple case is the \(\phi\)MDE phase, where using the specified values in the section \ref{sec:level42}, we obtain \(F = \left(\frac{1}{2}-3 \beta^2\right)\) and \(S = \frac{3}{2}\left(1+2\beta^2\right)\left(1-\frac{2}{3}\beta^2\right)\). Substituting these into Eq. \eqref{eq:f} yields the constant solution \(f = 1+2\beta^2\). This is particularly useful as the \(\phi\)MDE phase will serve as an initial point for further calculations. 

In CDE, the functions $F,S$, and therefore also $f$ if also the initial condition is $k$-independent, are independent of $k$. Below, we assume this to be the case in general.

\section{\label{sec:level3}Non-Linear Kernels}

We now move to higher orders in perturbations. We adopt the perturbation scheme of Eulerian standard perturbation theory. The general form of the conservation equations in the Newtonian limit and in Fourier space are (see e.g. the review in \cite{Bernardeau_2002}):

\begin{eqnarray}
 \label{eq:continuity}
\frac{\partial \delta(\mathbf{k}, \eta)}{\partial \eta}+\theta(\mathbf{k}, \eta)&&=-\int_{\mathbf{k} = \mathbf{k}_1+ \mathbf{k}_2}  
 \alpha\left(\mathbf{k}_1, \mathbf{k}_2\right) \theta\left(\mathbf{k}_1, \eta\right) \delta\left(\mathbf{k}_2, \eta\right) \, ,\\
 \label{eq:euler}
\frac{\partial \theta(\mathbf{k}, \eta)}{\partial \eta}+F\theta(\mathbf{k}, \eta)+ S\delta(\mathbf{k}, \eta)&&=-\int_{\mathbf{k} = \mathbf{k}_1+ \mathbf{k}_2} \beta\left(\mathbf{k}_1, \mathbf{k}_2\right)  \theta\left(\mathbf{k}_1, \eta\right)  \theta\left(\mathbf{k}_2, \eta\right) \, ,
\end{eqnarray}
where we defined:  \footnote{
The integral notation for general cases is defined as follows:
\begin{equation}
    \int_{\sum\mathbf{k}_{i}=\mathbf{k}}[...]=\int [\prod _i\frac{\mathrm{d}^3\mathbf{k}_i}{(2\pi)^3}] (2\pi)^3\delta_D \left( \sum_i\mathbf{k}_i-\mathbf{k} \right) [...]  \nonumber \,.
\end{equation}
 We use the following Fourier transform convention:
\begin{equation}
\tilde{f}(\mathbf{k}) = \int \mathrm{d}^3\mathbf{x} \, f(\mathbf{x})\, e^{-i \mathbf{k} \cdot \mathbf{x}}\,, \qquad
f(\mathbf{x}) = \int \frac{\mathrm{d}^3\mathbf{k}}{(2\pi)^3} \, \tilde{f}(\mathbf{k})\, e^{i \mathbf{k} \cdot \mathbf{x}} \nonumber \,,
\end{equation} 
such that the Dirac delta function satisfies
\begin{equation}
(2\pi)^3 \delta_D(\mathbf{k}) = \int \mathrm{d}^3\mathbf{x} \, e^{i \mathbf{k} \cdot \mathbf{x}} \nonumber \,.
\end{equation} }
\begin{eqnarray}  
\label{eq:3.29}
    \alpha\left(\mathbf{k}_1, \mathbf{k}_2\right) \equiv \frac{\left(\mathbf{k}_1+\mathbf{k}_2\right) \cdot \mathbf{k}_1}{k_1^2}, \quad \beta\left(\mathbf{k}_1, \mathbf{k}_2\right) \equiv \frac{\left|\mathbf{k}_1+\mathbf{k}_2\right|^2\left(\mathbf{k}_1 \cdot \mathbf{k}_2\right)}{2 k_1^2 k_2^2} \, ,\\
   \int_{\mathbf{k} = \mathbf{k}_1+ \mathbf{k}_2} = \int \frac{d^3 k_1}{(2\pi)^3} \frac{d^3 k_2}{(2\pi)^3}(2\pi)^3 \delta_{\mathrm{D}}\left(\mathbf{k}-\mathbf{k}_1-\mathbf{k}_2\right) \, .
\end{eqnarray}
The density contrast is defined  as $\delta= \frac{\rho(\eta)}{\bar{\rho}(\eta)} - 1$, with $\bar{\rho}(\eta)$ denoting the mean density. The term $\theta = \frac{i k_i v^i}{H a}$ describes the divergence of the peculiar velocity field in Fourier space.
 We consider the matter distribution to act as a pressureless fluid without vorticity. Additionally, we assume that the peculiar velocities are sufficiently small to be treated as non-relativistic and that we are 
at deep sub-horizon scales, $k \gg a H$, where the Newtonian approximations hold.

 To solve these equations, we need to apply perturbation theory, by expressing the solution as a functional of the linear matter density $\delta^{(1)}$. This approach may break down at very small scales due to additional stochastic effects from short-wavelength modes \cite{Nishimichi_2016,Baumann_2012,Carrasco_2012}; however,  in the weakly nonlinear regime considered here, these effects remain negligible. Consequently, we can expand the solution perturbatively \cite{Bernardeau_2002}, as shown below:
\begin{equation}
   \delta(\mathbf{\eta},\mathbf{k})=\sum_{n=1}^{\infty} \delta_{\mathbf{k}}^{(n)}(\eta), \quad \theta(\mathbf{\eta},\mathbf{k})=\sum_{n=1}^{\infty} \theta_{\mathbf{k}}^{(n)}(\eta) \, .
   \label{eq:pt}
\end{equation}
By inserting the expression \eqref{eq:pt} into the  conservation equations \eqref{eq:continuity},\eqref{eq:euler} and solving for each order separately, one can obtain  the following general form for {\it n}th order:
\begin{eqnarray}   
\delta_{\mathbf{k}}^{(n)}(\eta) & \equiv\frac{1}{n!} \mathcal{\int}_{\mathbf{k} = \mathbf{q}_1 +\cdots +\mathbf{q}_n} F_n\left(\mathbf{q}_1, \cdots, \mathbf{q}_n ; \eta\right) \delta^{(1)}_{\mathbf{q}_1}(\eta) \cdots \delta^{(1)}_{\mathbf{q}_n}(\eta) \, ,\label{eq:dens_pt} \\
\theta_{\mathbf{k}}^{(n)}(\eta) & \equiv \frac{1}{n!} f\mathcal{\int}_{\mathbf{k} = \mathbf{q}_1 +\cdots+ \mathbf{q}_n} G_n\left(\mathbf{q}_1, \cdots, \mathbf{q}_n ; \eta\right) \delta^{(1)}_{\mathbf{q}_1}(\eta) \cdots \delta^{(1)}_{\mathbf{q}_n}(\eta) \, .\label{eq:vel_pt} 
\end{eqnarray}

The \(\delta^{(1)}_{\mathbf{q}_n}(\eta)\) terms on the right-hand side represent linear density contrast , which evolve from the initial density  contrast via the growth function $D(\eta)$ as \(\delta^{(1)}_{\mathbf{q}_n}(\eta) = D(\eta) \delta^{(1)}_{\mathbf{q}_n}\). The functions \( F_n \) and \( G_n \), known as kernels, arise from nonlinear terms in the conservation equations, and they describe the coupling between different modes. For instance, in the Einstein-de Sitter Universe (EdS) model, the kernels can be derived recursively, starting from the linear order \cite{1994ApJ, Bernardeau_2002}. This form of the solution, as given in equations \eqref{eq:dens_pt} and \eqref{eq:vel_pt}, can also be applied to non-EdS models. However, determining the kernel structure for these models can be quite challenging. Therefore, we need to explore alternative methods.

We begin by deriving the equation for the linear-order growth  function $D(\eta)$. We neglect therefore the nonlinear terms on the right-hand sides (RHS) of equations~(\ref{eq:continuity}) and~(\ref{eq:euler}). Taking the time derivative of the continuity equation~(\ref{eq:continuity}), we obtain  $\delta''(\mathbf{k}, \eta) = -\theta'(\mathbf{k}, \eta)$.  We then substitute this result into the first term on the left-hand side (LHS) of the Euler equation~(\ref{eq:euler}). 
Finally, inserting the linear-order density contrast  $\delta^{(1)}_{\mathbf{k}}(\eta) = D(\eta)\, \delta^{(1)}_{\mathbf{k}}$,  
we  obtain a single equation for $D(\eta)$:

\begin{equation}
    D''+FD'- SD=0\,.
\end{equation}

In this work, we employ the perturbation theory kernel structure studied in Ref. \cite{D_Amico_2021}, which was obtained by using the symmetries of the conservation equations. This particular symmetry is extended Galilean invariance, where the equation of motion remains invariant under spatial translations with an arbitrary time parameter \cite{Kehagias_2013,Peloso_2013,1955ZA1,1956ZA2}. This approach is particularly advantageous for scenarios beyond $\Lambda$CDM. For instance, following the procedure outlined in \cite{D_Amico_2021}, one can express a generic second-order kernel as:  
\begin{equation}
    F_2\left(\mathbf{q}_1, \mathbf{q}_2 ; \eta\right)=a_0^{(2)}(\eta)+a_1^{(2)}(\eta) \gamma\left(\mathbf{q}_1, \mathbf{q}_2\right)+a_2^{(2)}(\eta) \beta\left(\mathbf{q}_1, \mathbf{q}_2\right) \, .
    \label{eq:F2}
\end{equation}

The \( F_2 \) matter kernel \eqref{eq:F2} is constructed from two basis functions, \( \gamma \) and \( \beta \), which depend on the external momenta \( \mathbf{q}_1 \) and \( \mathbf{q}_2 \). These functions are homogeneous, i.e. \( F(\lambda \mathbf{q}_1, \lambda \mathbf{q}_2) = F(\mathbf{q}_1, \mathbf{q}_2) \), and  symmetric under the exchange of \( \mathbf{q}_1 \) and \( \mathbf{q}_2 \), so that \( F(\mathbf{q}_1, \mathbf{q}_2) = F(\mathbf{q}_2, \mathbf{q}_1) \). Due to the isotropy of the Universe, the basis functions are also rotationally invariant. For external momenta \( \mathbf{p} \) and \( \mathbf{q} \), using rotational invariance and homogeneity, one can construct  basis functions as follows \cite{D_Amico_2021}
   \begin{equation}
    \text { 1, } \quad \gamma(\mathbf{q}, \mathbf{p})=1-\frac{(\mathbf{q} \cdot \mathbf{p})^2}{q^2 p^2}, \quad \beta(\mathbf{q}, \mathbf{p}) \equiv \frac{|\mathbf{q}+\mathbf{p}|^2 \mathbf{q} \cdot \mathbf{p}}{2 q^2 p^2}, \quad \alpha_a(\mathbf{q}, \mathbf{p})=\frac{\mathbf{q} \cdot \mathbf{p}}{q^2}-\frac{\mathbf{p} \cdot \mathbf{q}}{p^2} \, .
    \label{eq:basis}
 \end{equation}

Note that $\gamma$ and $\beta$  are  symmetric, while $\alpha_a$ is an anti-symmetric function. Here, \( a_0^{(2)} \), \( a_1^{(2)} \), and \( a_2^{(2)} \) are time-dependent coefficients, where \( a_0^{(2)} \) and \( a_2^{(2)} \)  can be determined by applying constraints derived from the symmetries of the  conservation equations. The remaining undetermined coefficient, \( a_1^{(2)} \), can be fixed by our cosmological model. The most general forms of the matter kernels (i.e density contrast kernel $F_i$ and the velocity divergence kernel $G_i$), after applying the constraints, are presented below 
\cite{D_Amico_2021}: 
\begin{eqnarray}  
F_1 && =1 \, ,\\
F_2\left(\mathbf{q}_1, \mathbf{q}_2;\eta\right) && =2 \beta\left(\mathbf{q}_1, \mathbf{q}_2\right)+a_1^{(2)} (\eta) \gamma\left(\mathbf{q}_1, \mathbf{q}_2\right) \, ,\\
F_3\left(\mathbf{q}_1, \mathbf{q}_2, \mathbf{q}_3;\eta\right) && =2 \beta\left(\mathbf{q}_1, \mathbf{q}_2\right) \beta\left(\mathbf{q}_{12}, \mathbf{q}_3\right)+a_5^{(3)} (\eta) \gamma\left(\mathbf{q}_1, \mathbf{q}_2\right) \gamma\left(\mathbf{q}_{12}, \mathbf{q}_3\right) -\\&&-2\left(a_{10}^{(3)}(\eta) -h(\eta) \right) \gamma\left(\mathbf{q}_1, \mathbf{q}_2\right) \beta\left(\mathbf{q}_{12}, \mathbf{q}_3\right)+ \nonumber \\ && 
+2\left(a_1^{(2)}(\eta)  +2 a_{10}^{(3)}(\eta) -h(\eta) \right) \beta\left(\mathbf{q}_1, \mathbf{q}_2\right) \gamma\left(\mathbf{q}_{12}, \mathbf{q}_3\right)+ \nonumber  \\ && +a_{10}^{(3)}(\eta)  \gamma\left(\mathbf{q}_1, \mathbf{q}_2\right) \alpha_a\left(\mathbf{q}_{12}, \mathbf{q}_3\right) +\text { cyclic } \, ,\nonumber
\end{eqnarray}
and
\begin{eqnarray}  
G_1&& =-1 \, , \\
G_2\left(\mathbf{q}_1, \mathbf{q}_2; \eta \right)   && =-2 \beta\left(\mathbf{q}_1, \mathbf{q}_2\right)-d_1^{(2)} (\eta) \gamma\left(\mathbf{q}_1, \mathbf{q}_2\right) \, , \\
 G_3\left(\mathbf{q}_1, \mathbf{q}_2, \mathbf{q}_3;\eta\right) && =-2 \beta\left(\mathbf{q}_1, \mathbf{q}_2\right) \beta\left(\mathbf{q}_{12}, \mathbf{q}_3\right)-d_5^{(3)} (\eta) \gamma\left(\mathbf{q}_1, \mathbf{q}_2\right) \gamma\left(\mathbf{q}_{12}, \mathbf{q}_3\right)+ \\
 && +2\left(d_{10}^{(3)}(\eta) -h(\eta) \right) \gamma\left(\mathbf{q}_1, \mathbf{q}_2\right) \nonumber \beta\left(\mathbf{q}_{12}, \mathbf{q}_3\right)-
 \\&&-2\left(d_1^{(2)}(\eta) +2 d_{10}^{(3)}(\eta) -h(\eta) \right) \beta\left(\mathbf{q}_1, \mathbf{q}_2\right) \gamma\left(\mathbf{q}_{12}, \mathbf{q}_3\right)- \nonumber\\
 &&-d_{10}^{(3)} (\eta) \gamma\left(\mathbf{q}_1, \mathbf{q}_2\right) \alpha_a\left(\mathbf{q}_{12}, \mathbf{q}_3\right)+\text { cyclic} \, .\nonumber
\end{eqnarray}

The $h$ function is defined as:
\begin{equation}
    h(\eta) \equiv \int^\eta d \eta^{\prime} f\left(\eta^{\prime}\right)\left[\frac{D\left(\eta^{\prime}\right)}{D(\eta)}\right]^2 d_1^{(2)}\left(\eta^{\prime}\right) \, ,
\end{equation}
and  \(\mathbf{q}_{ij \ldots} = \mathbf{q}_i + \mathbf{q}_j+\ldots\) is a shorthand notation for the sum of  momenta; for example,  \(\mathbf{q}_{12} = \mathbf{q}_1 + \mathbf{q}_2\).  At second order, $F_2$ depends on the single undetermined coefficient \( a^{(2)}_1 \), while $G_2$ depends on \( d^{(2)}_1 \). Moving to third order, \( F_3 \) includes three undetermined coefficients: \( a^{(2)}_1 \), \( a^{(3)}_5 \), and \(a^{(3)}_{10} \), whereas \( G_3 \) includes \( d^{(2)}_1 \), \( d^{(3)}_5 \), and \( d^{(3)}_{10}\).

\subsection{Solving the evolution equations for the third-order kernel}

In the previous section, we introduced the nonlinear kernel structure for the density contrast $F_i$ and velocity divergence $G_i$ fields up to third order. We now turn to deriving the evolution equations for the time-dependent kernel coefficients.  These equations have been given  in \cite{D_Amico_2021}, with an application   for the second order in both the $\Lambda$CDM and Gabadadze-Porrati (nGPD) models. Here  we  present a complete third-order derivation for the coupled dark energy model.  

\subsubsection{\label{sec:level311}Continuity equation}
We  express the explicit form of \eqref{eq:dens_pt} and \eqref{eq:vel_pt} at third order as follows:
\begin{equation}  
\delta^{(3)}_{k} (\eta)=\frac{1}{3!}  \int_{\mathbf{k} = \mathbf{k}_{123}} D\delta^{(1)}_{k_1}D \delta^{(1)}_{k_2} D\delta^{(1)}_{k_3} F_3(\mathbf{k}_1, \mathbf{k}_2,\mathbf{k}_3;\eta)  \, ,
\label{eq:dens3}
\end{equation}
\begin{equation}  
\theta^{(3)}_{k} (\eta) =\frac{1}{3!} f  \int_{\mathbf{k} = \mathbf{k}_{123}} D\delta^{(1)}_{k_1}D \delta^{(1)}_{k_2} D\delta^{(1)}_{k_3} G_3(\mathbf{k_1}, \mathbf{k}_2,\mathbf{k}_3;\eta)   \, .
\label{eq:vel3}
\end{equation}
 Inserting the expressions  \eqref{eq:dens3} and \eqref{eq:vel3} into the continuity Eq. \eqref{eq:continuity} and keeping only third order terms on the (RHS), we find: 
\begin{align}
\label{eq:3.31}
& \frac{1}{3!} \int_{\mathbf{k} = \mathbf{k}_{123}}
 \left( D^3 F_3 \delta^{(1)}_{k_1} \delta^{(1)}_{k_2} \delta^{(1)}_{k_3}  \right)' + \frac{1}{3!} \int_{\mathbf{k} = \mathbf{k}_{123}}   f D^3 G_3 \delta^{(1)}_{k_1} \delta^{(1)}_{k_2} \delta^{(1)}_{k_3}  =\\ &- \int_{\mathbf{k} = \mathbf{k}_{12} }
 \alpha\left(\mathbf{k}_1, \mathbf{k}_2\right) \left[ \theta^{(2)}\left(\mathbf{k}_1, \eta\right) \delta^{(1)}  \left(\mathbf{k}_2, \eta\right)+\theta^{(1)}\left(\mathbf{k}_1, \eta\right) \delta^{(2)} \left(\mathbf{k}_2, \eta\right) \right] \, .\nonumber
\end{align}
The common factor \(D^3\) cancels at the end and is omitted here. Expanding the second-order terms  on the (RHS) of Eq. \eqref{eq:3.31}, we obtain the following relation:
\begin{align}
 \label{eq:RHS1}
\text { RHS} \equiv  -\frac{1}{2!} f\int_{\mathbf{k} = \mathbf{k}_{12}} 
 \alpha\big(\mathbf{k}_1, \mathbf{k}_2\big) \bigg[
\int_{\mathbf{k_1} = \mathbf{q}_{12}} 
 G_2(\mathbf{q}_1, \mathbf{q}_2)F_1  \delta^{(1)}_{q_1}\delta^{(1)}_{q_2} \delta^{(1)}_{k_2}   +\\ +\int_{\mathbf{k_2} = \mathbf{q}_{12}} 
 F_2(\mathbf{q}_1, \mathbf{q}_2)G_1  \delta^{(1)}_{q_1}\delta^{(1)}_{q_2} \delta^{(1)}_{k_1}\bigg] \, . \nonumber
\end{align}
 To simplify, we contract the double integral in Eq. \eqref{eq:RHS1} as $\int_{\mathbf{k} =  \mathbf{q}_i+\mathbf{q}_j+\mathbf{k}_j}=\int_{\mathbf{k} = \mathbf{k}_i+\mathbf{k}_j}\int_{\mathbf{k_i} = \mathbf{q}_i+\mathbf{q}_j}$    and relabel the momenta ($\mathbf{q}_i \rightarrow \mathbf{k}_i, \mathbf{q}_j \rightarrow \mathbf{k}_k$ ;   \(i, j, k \in \{1, 2, 3\}\)). This results in: 
\begin{align}
\label{eq:RHS}
\text {RHS} \equiv -\frac{1}{2!}f\int_{\mathbf{k} ; \mathbf{k}_1, \mathbf{k}_2,\mathbf{k}_3} \bigg[\alpha\left(\mathbf{k}_{13}, \mathbf{k}_2\right)G_2(\mathbf{k}_1, \mathbf{k}_3)F_1  \delta^{(1)}_{k_1}\delta^{(1)}_{k_2} \delta^{(1)}_{k_3} + \\+ \alpha\left(\mathbf{k}_1,\mathbf{k}_{23} \right)
 F_2(\mathbf{k}_2, \mathbf{k}_3)G_1\delta^{(1)}_{k_1}\delta^{(1)}_{k_2} \delta^{(1)}_{k_3}\bigg] \, .\nonumber
\end{align}
We  express the interaction coefficient $\alpha$ in terms of the basis functions defined in  \eqref{eq:basis}  as follows:  
\begin{equation}
\alpha\left(\mathbf{k}_{jk}, \mathbf{k}_i\right)=(\gamma + \beta + \frac{\alpha_a}{2})_{jk,i} \, .
\label{eq:alpha}
\end{equation}

By using the symmetry under integration, we can cyclically (cyc) expand the (RHS)  \eqref{eq:RHS}. After symmetrization, this introduces two additional cycles, so we divide the (RHS) by 3. Then, substituting the definitions of  \(F_2\),  \(G_2\) and Eq. \eqref{eq:alpha}, and eliminating the integration terms \(\frac{1}{6}\int_{\mathbf{k}= \mathbf{k}_{123}}\) and \(\delta^{(1)}_{k_1}\delta^{(1)}_{k_2}\delta^{(1)}_{k_3}\) from both sides, we arrive at the following expression:
\begin{align}
   \text {RHS} \equiv f\bigg[ 4\gamma_{23,1}\beta_{2,3} + 4\beta_{23,1}\beta_{2,3} + \gamma_{23,1}\gamma_{2,3}(a^{(2)}_{1}+d^{(2)}_{1})+\beta_{23,1}\gamma_{2,3}(a^{(2)}_{1}+d^{(2)}_{1})  +\\ +\frac{{\alpha_a}_{1,23}\gamma_{2,3}}{2}(a^{(2)}_{1}-d^{(2)}_{1})+\text{cyc} \nonumber  \bigg] \, .
\end{align}
The left-hand side (LHS), on the other hand, can be written as follows:
\begin{equation}
\text {  LHS} \equiv 3fF_3 + (F_3)'+ fG_3 \, .
    \label{eq:LHS}
\end{equation}
Next, we insert the kernel expression into the (LHS)  Eq. \eqref{eq:LHS} and group the terms based on their basis functions. Given the independence of the basis functions, we can match terms on the (LHS) with those on the (RHS) corresponding to the same basis functions. This process will generate four equations, two of which are independent and the other two are linear combinations of these independent equations:
\begin{equation}
 \label{eq:a5}
\Big[3fa_5^{(3)}+(a_5^{(3)})'-fd_5^{(3)} \Big](\gamma_{1,2}\gamma_{12,3} +\text{cyc}) = f [a^{(2)}_{1}+d^{(2)}_{1} ](\gamma_{12,3}\gamma_{1,2} + \text{cyc}) \, ,
\end{equation}

\begin{align}
 \label{eq:hh}
\Big[-6f(a_{10}^{(3)}-h)-2((a_{10}^{(3)})'-h')&+2f(d_{10}^{(3)}-h)\Big]\times \\
&\times (\gamma_{1,2}\beta_{12,3}+\text { cyc}) =f\Big[a^{(2)}_{1}+d^{(2)}_{1}\Big](\beta_{12,3}\gamma_{1,2}+\text{cyc}) \, ,\nonumber
\end{align}

\begin{align}
\label{eq:hhh}
\Big[6f(a_1^{(2)}+2 a_{10}^{(3)}-h)+2((a_1^{(2)})'+2( a_{10}^{(3)})'&-h')-2f(d_1^{(2)}+2 d_{10}^{(3)}-h)\Big] \times\\ &\times (\beta_{1,2}\gamma_{12,3}+\text{cyc})  =  4f( \gamma_{12,3}\beta_{1,2} + \text {cyc}) \, ,\nonumber
\end{align}

\begin{equation}
\label{eq:a10}
\Big[3fa_{10}^{(3)}+ (a_{10}^{(3)})'-fd_{10}^{(3)}\Big](\gamma_{1,2} {\alpha_a}_{3,12}+\text { cyc}) =-f\frac{\Big[a^{(2)}_{1}-d^{(2)}_{1}\Big]}{2}({\alpha_a}_{1,23}\gamma_{2,3}+\text{cyc}) \, .
\end{equation}

We select equations \eqref{eq:a5} and \eqref{eq:a10} , and after eliminating the basis functions from both sides, we present them in their final form as equations \eqref{eq:a5f} and \eqref{eq:a10f}.

\subsubsection{\label{sec:level32}Euler equation}

 We can proceed to  derive the remaining equations  using the Euler Eq. \eqref{eq:euler}. By following the same steps as outlined in the previous section, we obtain:
\begin{equation}
    f'G_3+3f^2G_3 + fG'_3+FfG_3+SF_3=-2f^2\left[\beta_{13,2}G_2(k_1,k_3)G_1+\text {cyc}\right]  \, .\label{eq:euler2}
\end{equation}  
To simplify the expression, we add  and subtract $SG_3$ from the (LHS) Eq. \eqref{eq:euler2}. 
\begin{equation}
  \text { LHS} \equiv 2f^2G_3+ \left(f'+f^2+Ff-S\right)G_3+fG'_3+S\left(G_3+F_3\right) \, .
    \label{eq:LHSeuler}
\end{equation} 
Now, we focus on the second term in \eqref{eq:LHSeuler}
. This term corresponds to the growth rate equation \eqref{eq:f}, which allows us to cancel it out. After dividing both sides by $f^2$, we find:
\begin{equation}
2G_3+\frac{G'_3}{f}+\frac{S}{f^2}(G_3+F_3) =\left(-4\beta_{13,2}\beta_{1,3}-2d^{(2)}_{1} \gamma_{1,3}\beta_{13,2}+\text {cyc}\right) \, .
\label{eq:euler3}
\end{equation}

As we have done previously, we substitute the kernels   expression into Eq. \eqref{eq:euler3} and solve for each basis function separately. This results in a set of equations, of which we present the two independent ones below:
\begin{eqnarray}
\left[-2d_{5}^{(3)}-\frac{(d_{5}^{(3)})'}{f}+\frac{S}{f^2}(a_{5}^{(3)}-d_{5}^{(3)})\right](\gamma_{1,2}\gamma_{12,3} +\text {cyc})=0 \, ,
\label{eq:d5}\\
\left[-2d_{10}^{(3)}-\frac{(d_{10}^{(3)})'}{f}+\frac{S}{f^2}(a_{10}^{(3)}-d_{10}^{(3)})\right](\gamma_{1,2} {\alpha_a}_{3,12} + \text {cyc})=0 \, .
\label{eq:d10}
\end{eqnarray}
After eliminating the basis functions from Eqs. \eqref{eq:d5} and \eqref{eq:d10}, we obtain the final form \eqref{eq:d5f} and \eqref{eq:d10f}. 
 \subsubsection{\label{sec:level33}Evolution equations for the kernel coefficients}

Using the kernel forms within the conservation equations, we derived the following system of coupled differential equations \cite{D_Amico_2021}:

\begin{eqnarray} 
\label{eq:a1f}
(a_1^{(2)})'&&=f(2-2a_1^{(2)}+d_1^{(2)}) \, ,\\
\label{eq:d1f}
(d_1^{(2)})'&&=-fd_1^{(2)}+\frac{S}{f}\left(a_1^{(2)}-d_1^{(2)}\right) \, ,
\\
\label{eq:a5f}
(a_5^{(3)})'&&=f\left(a_1^{(2)}+d_1^{(2)}-3a_5^{(3)}+d_5^{(3)}\right) \, , \\
\label{eq:d5f}
(d_5^{(3)})'&&=-2fd_5^{(3)}+ \frac{S}{f}\left(a_5^{(3)}-d_5^{(3)}\right) \, ,
\\
\label{eq:a10f}
(a_{10}^{(3)})'&&=-f\frac{1}{2}\left(a_1^{(2)}
-d_1^{(2)}+6a_{10}^{(3)}-2d_{10}^{(3)}\right) \, , \\
\label{eq:d10f}
(d_{10}^{(3)})'&&=-2fd_{10}^{(3)}+\frac{S}{f}\left(a_{10}^{(3)}-d_{10}^{(3)}\right) \, .
\end{eqnarray}

 These equations are general (provided that $S,f$ are $k$-independent) and applicable to various cosmological models specified through the determination of the source terms \( S \) and the growth rate \( f \). The source terms S, which depend on the specific cosmological model, enter only through the Euler equation. At third order, we obtained  Eq. \eqref{eq:a5f} and  \eqref{eq:a10f}  from the continuity equation, and Eq. \eqref{eq:d5f} and  \eqref{eq:d10f} from the Euler equation. At second order, we derived Eq. \eqref{eq:a1f}  from the continuity equation, and Eq. \eqref{eq:d1f} from the Euler equation. This results in six equations that describe the evolution of the time-dependent kernel coefficients.
\section{\label{sec:level4}Kernels for Coupled Dark Energy}

We are finally in place to express the kernels for our coupled dark energy model. After recalling the standard result for EdS, we examine explicitly three cases: 1) the $\phi$MDE epoch; 2) a linear potential; 3) an exponential potential. In the first case  we can provide analytical solutions, while in the others we need to perform numerical integrations.

\subsection{\label{sec:level41}Einstein-de Sitter}
We begin by calculating the kernel coefficients for the  Einstein-de Sitter (EdS) model, which represents a flat, matter-dominated universe.
The EdS universe is characterized by $x=0,y=0,z=0$,  $\Omega_m=1$ and $f=1$, which gives  $S=\frac{3}{2}$. Substituting these values into equations (\ref{eq:a1f}-\ref{eq:d10f}) and solving them, we find the following constant solutions \cite{D_Amico_2021}:
\begin{equation}
a^{(2)}_{1}=\frac{10}{7}, \quad d^{(2)}_{1}=\frac{6}{7}, \quad  a^{(3)}_{5}=\frac{8}{9}, \quad 
 d^{(3)}_{5}=\frac{8}{21},  \quad  a^{(3)}_{10}=-\frac{1}{9}, \quad  d^{(3)}_{10}=-\frac{1}{21} \, .
 \end{equation}
More generally, these solutions can also be obtained using recursion relations \cite{Goroff,Bernardeau_2002}.

\subsection{\texorpdfstring{$\phi$MDE}{phiMDE}\label{sec:level42}}

In this section, we briefly outline how we obtain the \(\phi\)-Matter-Dominated Epoch (\(\phi\)MDE) and calculate its kernel coefficients. From this point onward, we  ignore radiation and baryons due to their negligible contributions.

 The trajectories of solutions for \(x\) and \(y\) exist in the phase space, confined to the upper half of the unit disk, defined by \(x^2 + y^2 < 1\) and \(y \geq 0\). To determine the critical points, we set \(x' = y' = 0\) and solve equations \eqref{eq:x},\eqref{eq:y}. This  yields five critical points, which can be classified as stable, unstable, or saddle points based on their behavior in the phase space.

 In the uncoupled case where \( \beta = 0 \), the origin of the phase space, (\( x = y = 0 \)), is a critical point corresponding to a matter-dominated universe \( \Omega_m = 1 \). When a non-zero coupling parameter \(\beta\) is introduced, this critical point shifts from the origin along the \(x\)-axis. Now, it no longer characterizes a pure matter-dominated era; instead, due to contributions from the scalar field, we refer to it as the \(\phi\)-Matter-Dominated Epoch (\(\phi\)MDE). This epoch acts as a saddle point if $\beta< \sqrt{3/2}$ and is a viable matter era if $\beta\ll 1$. It is therefore a transient solution between the radiation-dominated era and the  late-time acceleration.

In the $\phi$MDE phase, the parameters are given by \( x = \sqrt{\frac{2}{3}} \beta \) and \( y = 0 \). This leads to the following expressions: 
$\Omega_K=\frac{2}{3}\beta^2, \Omega_P=0$,
\(\Omega_c = 1 - \frac{2}{3}  \beta^2\), \(S = \frac{3}{2} \left(1 - \frac{2}{3} \beta^2\right) \left(1 + 2 \beta^2\right)\),  \(f = 1 + 2 \beta^2\), and $w_{\rm eff} = \frac{2}{3}\beta^2$. Substituting these parameter values into equations (\ref{eq:a1f}-\ref{eq:d10f}), we obtain: 
\begin{align}
\label{eq:phiMDEcoeff}
a^{(2)}_{1}= \frac{4 \beta ^2+10}{6 \beta ^2+7} ,  \quad d^{(2)}_{1}= \frac{6-4 \beta ^2}{6 \beta ^2+7} , &\quad a^{(3)}_{5}=\frac{8}{10 \beta ^2+9} ,\\ \quad d^{(3)}_{5}= \frac{24-16 \beta ^2}{60 \beta ^4+124 \beta ^2+63} , \quad a^{(3)}_{10}= -\frac{2 \beta ^2+1}{10 \beta ^2+9} , & \quad d^{(3)}_{10}= \frac{4 \beta ^4-4 \beta ^2-3}{60 \beta ^4+124 \beta ^2+63}  \, . \nonumber 
\end{align}
As expected, for $\beta=0$, the solutions \eqref{eq:phiMDEcoeff} reduce to the EdS form. 

Since the evolution of CDE passes through $\phi$MDE regardless of the potential, we need to use the solutions above as initial value when solving numerically for the subsequent evolution. For this purpose, we set the initial time to $\eta = -3$  (corresponding to a redshift of approximately 19)  to calculate the $x$ and $y$ parameters. The potential-to-kinetic energy ratio is $\frac{\Omega_{Pi}}{\Omega_{Ki}} = 0.04$ at this point, confirming that the system is well within  $\phi$MDE. Figure \ref{fig:energy-frac-log} illustrates how these energy fractions evolve, starting from the $\phi$MDE phase.

\subsection{\label{sec:level44}Linear potential}

This section presents the analytical solutions for the parameters $x$ and $y$ for a scalar field with a linear potential.
This case is interesting for two reasons. First, if $\mu=0$, we have what can be perhaps defined a minimal modified gravity model, i.e. a dark-dark coupling plus a cosmological constant. This model introduces a single additional parameter to $\Lambda$CDM to fully describe background and perturbations. Secondly, the background behavior can be solved analytically if we assume that it deviates only slightly from $\Lambda$CDM, i.e. $\beta,\mu \ll 1$.

For the  linear potential $V \approx V_0 - V_0\mu\phi
$ the Klein-Gordon equation is:
\begin{equation}
    {\phi}''+(3+\frac{H'}{H}){\phi}'-  \frac{V_{0} \mu}{H^2}= 3\beta \Omega_{c} \, .
    \label{eq:linearKG}
\end{equation}

In our calculations, radiation and baryons are ignored, resulting in \(\Omega_m = \Omega_c\)    and for a flat universe we have,  $\Omega_{\Lambda 0} = 1 - \Omega_{m0}$. Next, taking derivative of \eqref{eq:H/H'} and using \eqref{eq:matterdens} and \eqref{eq:scalar_field_dens},  we  obtain:
\begin{equation}\frac{H'}{H}=\frac{-3(\Omega_{m}+\Omega_{\phi}+ \Omega_{\phi}w_{\phi})}{2} \, .
\end{equation}

We now assume that \( w_{\phi} \) is nearly \( -1 \), which allows us to approximate \( \frac{H'}{H} \approx\frac{-3\Omega_{m}}{2} \). This assumption is valid for our model, which slightly deviates from \(\Lambda\text{CDM}\) and is  accurate for current values. Accordingly, with \( w_{\phi} = -1 \), $\Omega_{m}$ can be expressed as $\frac{\Omega_{m0}}{\Omega_{m0} + \Omega_{\Lambda0} e^{3\eta}}$, with \( \Omega_{m0} \) and \( \Omega_{\Lambda0} \) representing the present matter density fraction and present scalar field density fraction, respectively. Thus, using these expressions in Eq. \eqref{eq:linearKG}, we obtain the following analytical  solution for \( x = \frac{\phi'}{\sqrt{6}} \):

\begin{align}
\label{eq:linearx}
x=\frac{1}{\sqrt{6}\,e^{3\eta}E(\eta)}
\Bigg\{
&2\beta\left[
e^{3\eta_i}E(\eta_i)
+\frac{\Omega_{m0}}{\sqrt{\Omega_{\Lambda 0}}}
\ln\!\left(
\frac{e^{3\eta_i/2}(E(\eta_i)-\sqrt{\Omega_{\Lambda 0}})}
{e^{3\eta/2}(E(\eta)-\sqrt{\Omega_{\Lambda 0}})}
\right)
\right] \\&
+ \mu\left[
\frac{\Omega_{m0}}{\sqrt{\Omega_{\Lambda0}}}
\ln\!\left(
\frac{e^{3\eta/2}(E(\eta)-\sqrt{\Omega_{\Lambda 0}})}
{e^{3\eta_i/2}(E(\eta_i)-\sqrt{\Omega_{\Lambda 0}})}
\right)
+e^{3\eta}E(\eta)-e^{3\eta_i}E(\eta_i)
\right]\nonumber
\Bigg\} \, ,
\end{align}
where \( E(\eta) \) is defined as \( \sqrt{\Omega_{m0} e^{-3\eta} + \Omega_{\Lambda0}} \), with  \( \eta_{i} \) indicating the initial time. We choose the initial time to be well within the $\phi $MDE, as discussed in  Sec. \ref{sec:level42}. When $\mu = 0$, indicating constant potential, the solution reduces to the first term in Eq. \eqref{eq:linearx}. This term  depends linearly on the coupling strength, whereas the second term is unaffected by the coupling.

\begin{figure}
        \centering
     \includegraphics[width=0.55\textwidth]{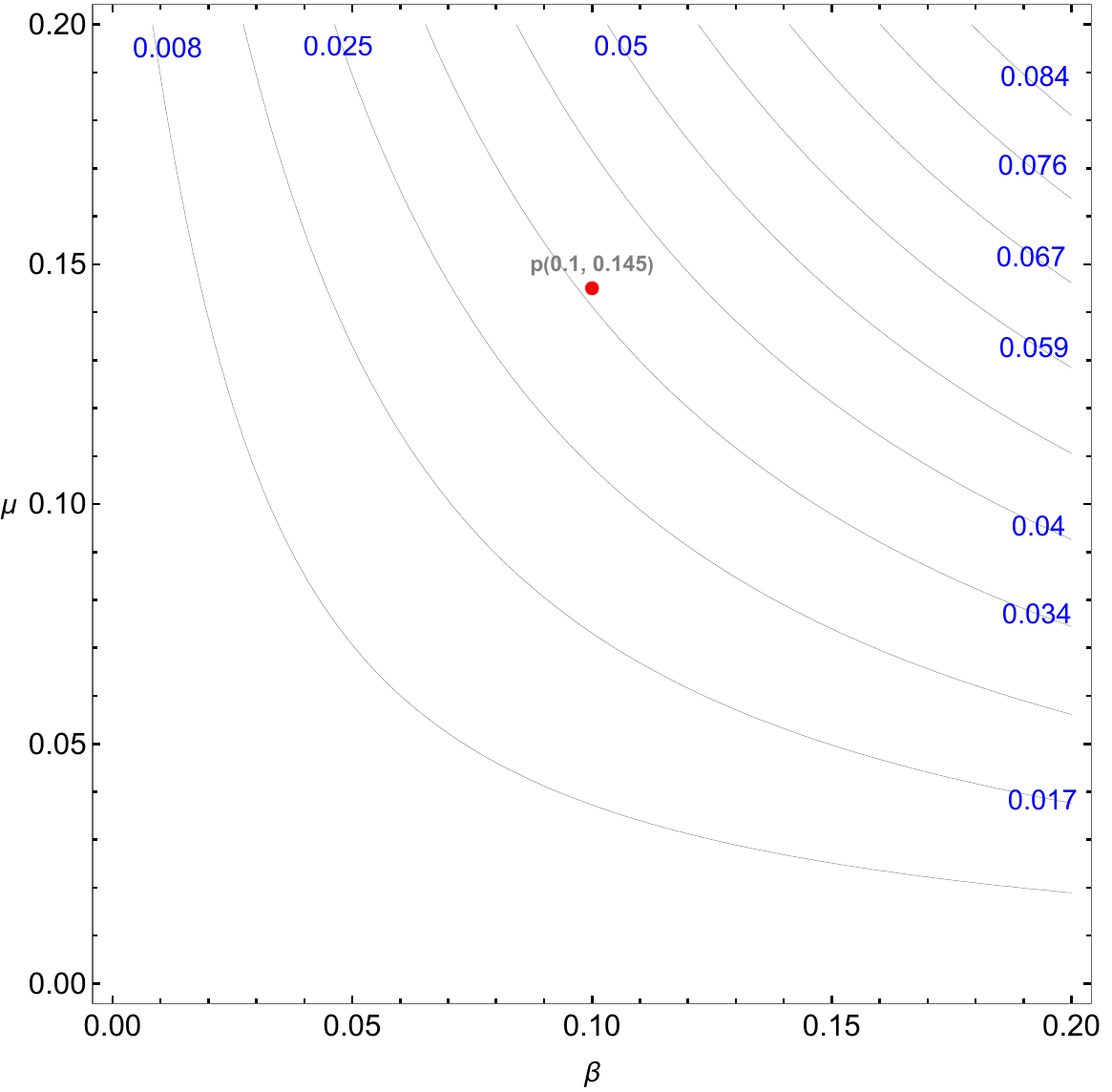}
        \caption{ This figure displays the values of $p(\beta, \mu) = \mu \int \phi'(\eta,\beta) \, d\eta$ on the contour, which allows us to determine the parameter values of \(\beta\) and $\mu$  that satisfy the linearity condition $p \ll 1$. The red point represents the reference values \(\beta = 0.1\) and \(\mu = 0.145\). Contour levels  are labeled in blue.} 
        \label{fig:linear_cond}
  \end{figure}

\begin{figure}
    \centering
    \begin{subfigure}[b]{0.46\textwidth}
        \centering
        \includegraphics[width=\textwidth]{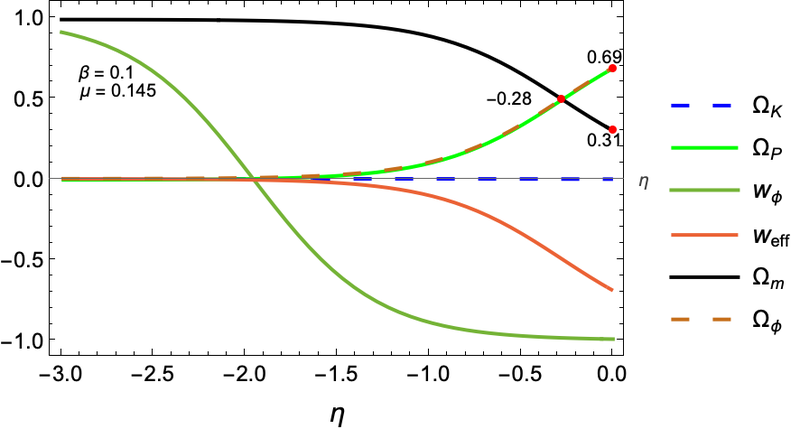}
        \caption{}
        \label{fig:energy-frac}
    \end{subfigure}
    \hspace{0.05\textwidth} 
    \begin{subfigure}[b]{0.46\textwidth}
        \centering   \includegraphics[width=\textwidth]{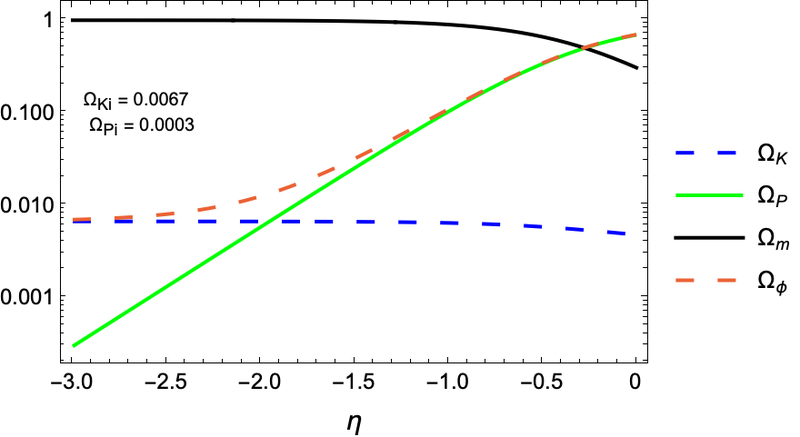}
        \caption{}
        \label{fig:energy-frac-log}
    \end{subfigure}
    
 \caption{ Figure (a) shows the background parameters — energy fractions and state parameters — based on analytical solutions of \(x\) and \(y\) for a linear potential. Figure (b) displays these energy fractions in logarithmic scale. $\Omega_{Pi}$ and $\Omega_{Ki}$ correspond to the initial values of potential and kinetic energy, respectively.
} 
\end{figure}

To determine the range of $\mu$ and $\eta$ where the linear approximation $\mu \Delta \phi \ll 1$ holds,  we can define $p(\beta, \mu) = \mu \int \phi'(\eta,\beta) \, d\eta$   and carry out the integration over the  range $-3 < \eta < 0$.  This results in a contour plot for $p(\beta,\mu)$ shown in Figure \ref{fig:linear_cond}. In the following, we will use as reference values $\mu=0.145$,  $\beta=0.1$ (marked with a red dot in the figure \ref{fig:linear_cond}), for which the linearity condition is well verified. In Fig. (\ref{fig:energy-frac}), we show that for this choice of parameters, the EoS $w$ is approximately -1 today, confirming that the background is close to $\Lambda$CDM.

We can find 
$y(\eta)=\frac{1}{H_{0}E(\eta)}\sqrt{\frac{V(\eta)}{3 }}$, by  solving for $V(\eta)$  using the relation $V'(\eta )=-\mu V_0 {\phi}'=-\mu\sqrt{6}V_0 x(\eta )$. This gives the following result:

\begin{align}
V = V_0 \Bigg\{
1 - 2 \beta \eta \, \mu 
+&\mu \Bigg[
\frac{4\beta-2\mu}{3\sqrt{\Omega_{\Lambda0}}}
\Bigg(
\frac{\sqrt{\Omega_{\Lambda0}}}{\Omega_{m0}} e^{3\eta_i} E(\eta_i) (E(\eta)-1)
\\&- \ln\!\Big( \frac{e^{3\eta_i/2}(E(\eta_i)-\sqrt{\Omega_{\Lambda0}})}{1-\sqrt{\Omega_{\Lambda0}}} \Big)
+ E(\eta) \ln\!\Big( \frac{e^{3\eta_i/2}(E(\eta_i)-\sqrt{\Omega_{\Lambda0}})}{e^{3\eta/2}(E(\eta)-\sqrt{\Omega_{\Lambda0}})} \Big)
\Bigg)
\Bigg]
\Bigg\} \, . \nonumber
\end{align}

\begin{figure}
 \centering
    \begin{subfigure}[b]{0.46\textwidth}
      
        \includegraphics[width=\textwidth]{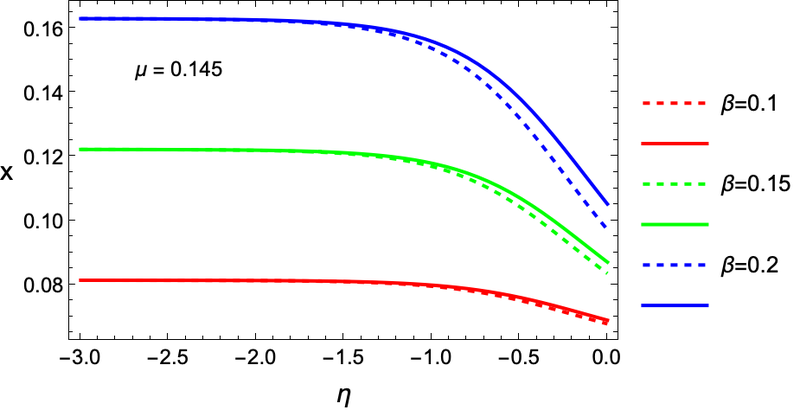}
        \caption{}
    \end{subfigure}
    \hspace{0.05\textwidth} 
    \begin{subfigure}[b]{0.46\textwidth}
     
        \includegraphics[width=\textwidth]{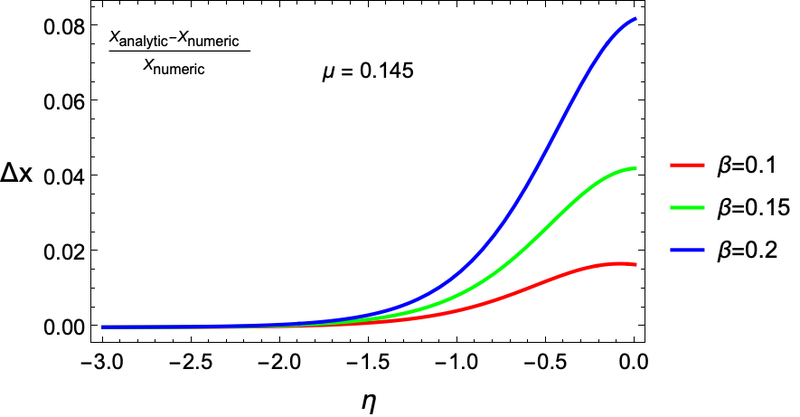}
      \caption{}
    \end{subfigure}

    \begin{subfigure}[b]{0.46\textwidth}
        \includegraphics[width=\textwidth]{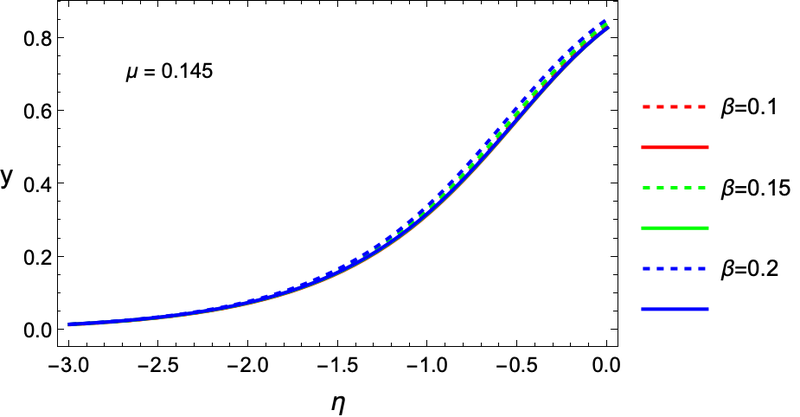} 
        \caption{}
    \end{subfigure}
    \hspace{0.05\textwidth} 
    \begin{subfigure}[b]{0.46\textwidth}
        \includegraphics[width=\textwidth]{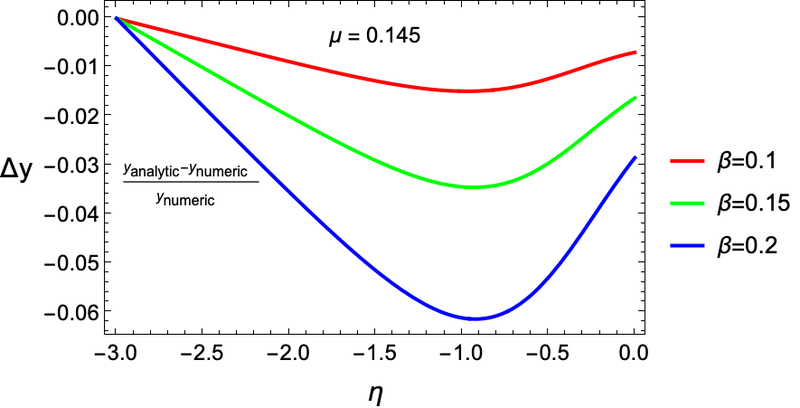}
        \caption{}
    \end{subfigure}
    \caption{
     Figures {\it a} and {\it c} compare the analytical (solid line) and numerical (dashed line) solutions for $x$ and $y$, respectively. Figures {\it b} and {\it d} illustrate the relative deviation of analytic from numeric ($\frac{\text{analytic}}{\text{numeric}}-1$)   for $x$ and $y$, showing that the deviation increases with larger values of \(\beta\). }
     \label{fig:num_analy}
\end{figure}

After calculating $x(\eta)$ and $y(\eta)$ analytically, we solve the equation for $f(\eta)$ and the evolution equations of the coefficients (\ref{eq:a1f}-\ref{eq:d10f}) numerically (see figures \ref{fig:num_analy},\ref{fig:coeff/phimde}, and \ref{fig:growth_rate} for a comparison of the analytical and numerical solutions).  
We present the fitting functions for these numerical solutions of kernel coefficients for the linear and exponential potential cases in the appendix \ref{Appendix:Fitting functions}. The functions are precise to within 1\% or better in the range $\beta\in (0,0.2),\mu\in (0,0.2)$ and $\eta\in (-1.4,0)$. This range corresponds to the  redshift range that can be observed in the near future and to values of the parameters that are close to the current constraints. We estimate that within these ranges, our functions may differ from  EdS values by up to 2\% for \(a_{1}^{(2)}\), 6\% for \(d_{1}^{(2)}\), 4\% for \(a_{5}^{(3)}\), 10\% for \(d_{5}^{(3)}\), 3\% for \(a_{10}^{(3)}\), and 4\% for \(d_{10}^{(3)}\). These deviations appear large enough to be detected in forthcoming surveys. Therefore, we expect they can improve future constraints on the CDE parameters $\beta$ and $\mu$.

\section{\label{sec:level6}Conclusions}

In this paper, we studied non-linear corrections to one-loop of a coupled dark energy model (CDE), characterized by a dark-dark coupling $\beta$ and a potential with linear or exponential slope $\mu$. The linear case is meant to be an approximation to a generic potential with a sufficiently flat slope. The case of zero slope, in which the potential reduces to a cosmological constant, can be seen as a minimal modified gravity model with just a single parameter beyond $\Lambda$CDM.

The CDE model affects simultaneously the background evolution, the linear growth, the non-linear kernel coefficients, and the initial conditions.
We provide analytical and numerical solutions for all these quantities, with  precise fitting functions in the relevant range. 

These CDE kernels can be directly inserted into the general expressions for the power spectrum and bispectrum that apply to tracers in redshift space like galaxies (e.g. \cite{2015JCAP,Desjacques:2016bnm}). In this form, they are suitable for a comparison with real data that will be produced by ongoing and future surveys. This will be the goal of further work.

\section*{Acknowledgments}
LA acknowledges support by the Deutsche Forschungsgemeinschaft (DFG, German Research Foundation) under Germany's Excellence Strategy EXC 2181/1 - 390900948 (the Heidelberg STRUCTURES Excellence Cluster).

\appendix
\section{\label{Appendix:Fitting functions}Appendix: Fitting Functions}

This appendix presents the fitting functions of  kernel coefficients and of the growth rate $f$ as functions of the parameters \(\eta\), \(\beta\), and \(\mu\), for the linear and exponential potentials. For the linear potential, we derived the \(x\) and \(y\) functions analytically, while the \(f\) and kernel coefficients were computed numerically. On the other hand, all calculations for the exponential potential were carried out numerically. To construct the fitting functions, we used polynomial basis functions, with \(\beta\) and \(\mu\) ranging from \(0\) to \(0.2\) with a step size of \(0.025\), and \(\eta\) ranging from \(-1.4\) to \(0\) with the same step size. While our starting point for the numeric solution is $\eta = -3$,  we truncate the fitting functions at $\eta = -1.4$ (corresponding to a redshift of approximately 3)  to improve the precision of the fit and also to cover the span of most current and future LSS surveys.

The initial value for \(x\) is chosen to lie on the $\phi$MDE, while  \(y\) is adjusted to match the current values  (\(\Omega_{\Lambda 0} = 0.69\), \(\Omega_{m0} = 0.31\) at $\beta=0$ and $\mu=0$), with the following values:
\[
x(-3) = \sqrt{\frac{2}{3}}\beta, \quad y(-3) = 0.0166 \, .
\]
In addition,  the initial conditions for the kernel coefficients are set to correspond to the \(\phi\)MDE coefficients \eqref{eq:phiMDEcoeff}.

To evaluate the accuracy of the fitting functions, we include a table of Relative Root Mean Square Error (RRMSE) and Relative  Maximum Absolute Deviation (RMAD). The definitions of these statistical measures are provided below:
\begin{equation*}
{\rm RRMSE} = \sqrt{\frac{1}{n} \sum_{i=1}^{n} (\frac{y_i - \hat{y}_i}{y_i})^2}\, , \quad {\rm RMAD}= {\rm Max}({\rm Abs}(\frac{y_i - \hat{y}_i}{y_i})) \, ,
\end{equation*}
where \(y_i\) are the data points, \(\hat{y}_i\) are the predicted values from the fitting function, and \(n\) is the number of data points.
We aim at RMAD better than 1\% and RRMSE better than 0.2\% across the above mentioned range.

 We find that  the  linear potential  coefficients may deviate from the  exponential potential ones by up to 0.1\% for \(a_{1}^{(2)}\), 1\% for \(d_{1}^{(2)}\), 0.1\% for \(a_{5}^{(3)}\), 3\% for \(d_{5}^{(3)}\), 0.2\% for \(a_{10}^{(3)}\), and 2\% for \(d_{10}^{(3)}\). This confirms that the linear potential is a good approximation to the exponential one.

\subsection*{Linear potential}

\scalebox{0.85}{
\hspace*{-0.4cm}

\renewcommand{\arraystretch}{1.5}
\begin{tabular}{|c|p{13cm}|c|c|} \hline 
 \multicolumn{2}{|c|}{Fitting Functions }& RRMSE&RMAD\\ \hline

$a_{1}^{2}$ &  \(\left(-0.2418 \beta ^2-0.0038 \beta +0.0132 \eta ^2-0.1070 \eta -0.0057 \mu +1.4324\right) \newline \times \left(-0.2536 \beta ^2+0.0796 \eta +0.0033 \mu +1\right)
\)  & 0.00004&    0.0002\\ \hline  

$d_{1}^{2}$ & 
\(
 \left(-0.5178 \beta ^2-0.0069 \beta +0.0425 \eta ^2-0.0976 \eta -0.0161 \mu +0.8772\right) \newline \times \left(-0.7925 \beta ^2+0.1558 \eta +0.0127 \mu +1\right)
\)  & 0.0002&0.0020\\ \hline 

$a_{5}^{3}$ & 
\(\left(
-0.9037 \beta ^2-0.0078 \beta +0.0040 \eta ^2+0.0096 \eta -0.0006 \mu +0.8950\right)
\)  & 0.0002&0.0013\\ \hline 

$d_{5}^{3}$ & 
\(\left(-0.3799 \beta ^2-0.0170 \beta -0.0053 \eta ^2+0.0636 \eta +0.0103 \mu +0.3964\right)\newline \times
\left(-15.5058 \beta ^3+0.0736 \eta ^2-0.0347 \mu +1\right)\newline \times\left(55.082 \beta ^4-0.0874 \eta +1\right)
\)  & 0.0004&0.0020\\ \hline

$a_{10}^{3}$ & 
\(\left(
-0.0898 \beta ^2-0.0008 \beta +0.0006 \eta ^2+0.0016 \eta -0.0001 \mu -0.1101\right)
\)  & 0.0002&0.002\\ \hline

$d_{10}^{3}$ & 
\(
\left(0.0218 \beta ^2-0.0002 \beta -0.0004 \eta ^2-0.0104 \eta -0.0006 \mu -0.0492\right) \newline 
\times\left(-0.1419 \beta ^2+0.0692 \eta ^2-0.1469 \eta -0.0213 \mu +1\right) 
\)  & 0.0003&0.002\\ \hline 

$f$ & 
\(\left(0.1425 \log \left(-0.8016 \beta  \mu +3.1240 \eta ^2-1.7161 \eta +1\right)+0.1893\right) \newline \times
\left(1.7700 e^{2.5303 \beta ^2+0.1534 \eta }+1\right) \times
\left(0.1277 \beta ^{3/2}+0.1494 \eta +1\right) 
\)  & 0.0015& 0.009\\ \hline
\end{tabular}
}

\subsection*{Exponential potential}

\scalebox{0.85}{
\hspace*{-0.4cm}

\renewcommand{\arraystretch}{1.5}
\begin{tabular}{|c|p{13cm}|c|c|}\hline
 \multicolumn{2}{|c|}{Fitting Functions}& RRMSE&RMAD\\
\hline
$a_{1}^{2}$ & 
\(
\left(-0.2548 \beta ^2-0.004937 \beta +0.01469 \eta ^2-0.1148 \eta -0.01117 \mu +1.432\right) \newline \times \left(-0.2432 \beta ^2+0.08508 \eta +0.006371 \mu +1\right)
\)  & 0.0001&0.0003\\ 
\hline
$d_{1}^{2}$ & 
\(
\left(-0.5648 \beta ^2-0.01006 \beta +0.04168 \eta ^2-0.09732 \eta -0.03224 \mu +0.8776\right)\newline \times \left(-0.7460 \beta ^2+0.1551 \eta +0.02543 \mu +1\right)
\)  & 0.0003&0.0028\\
\hline
$a_{5}^{3}$ & 
\(
-0.9046 \beta ^2-0.008510 \beta +0.003811 \eta ^2+0.009371 \eta -0.0007639 \mu +0.8950
\)  & 0.0002&0.0014\\
\hline
$d_{5}^{3}$ & 
\( 
 \left(-0.3703 \beta ^2-0.01909 \beta -0.005780 \eta ^2+0.05841 \eta +0.02074 \mu +0.3965\right)\newline \times\left(-16.11 \beta ^3+0.06935 \eta ^2-0.06856 \mu +1\right)  \newline \times\left(57.91 \beta ^4-0.07615 \eta +1\right) 
\)  & 0.0006&0.0033\\
\hline
$a_{10}^{3}$ & 
\(
\left(-0.08981 \beta ^2-0.0009346 \beta +0.0006462 \eta ^2+0.001532 \eta -0.0002591 \mu -0.1101 \right)
\)  & 0.0002&0.0017\\
\hline

$d_{10}^{3}$ & 
\(
\left(0.02688 \beta ^2+0.00004414 \beta +0.001978 \eta ^2-0.002243 \eta -0.002563 \mu -0.04920\right) \newline \times \left(-0.05780 \beta ^2+0.07670 \eta ^2+0.01473 \eta -0.06463 \mu +1\right) 
\)  & 0.0005&0.0047\\
\hline
$f$ & 
\(
\left(0.189 \log (8.419 \beta  \mu -1.403 \beta +5.155 \eta ^2+1)+0.7806\right) \times \left(1-0.3277 e^{2.897 \eta }\right) \newline \times \left(0.738 \beta ^{3/2}+0.1469 \eta +1\right)
\)  & 0.0024&0.0084\\
\hline

\end{tabular}

}

\begin{figure}
    \centering
    \begin{subfigure}[b]{0.46\textwidth}
        \centering
        \includegraphics[width=\textwidth]{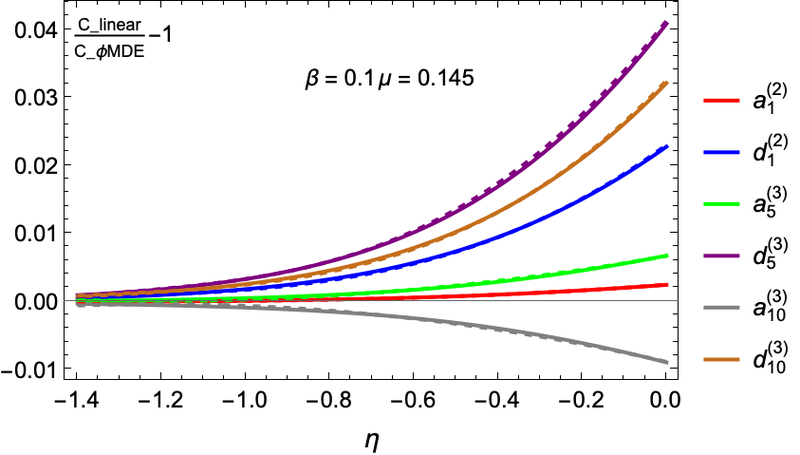}
        \caption{}
    \end{subfigure}
   \hspace{0.05\textwidth}
    \begin{subfigure}[b]{0.46\textwidth}
        \centering
        \includegraphics[width=\textwidth]{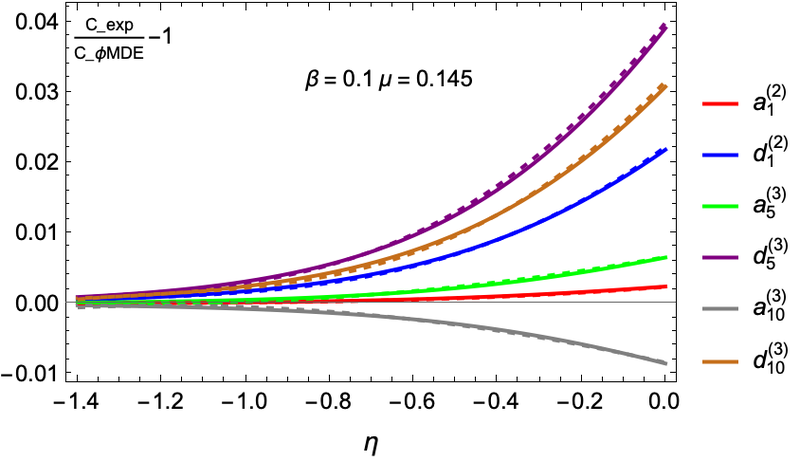}
        \caption{}
    \end{subfigure}
   \caption{These figures illustrate the relative deviation of kernel coefficients from the \(\phi\)MDE case, expressed as \(\frac{C}{C_{\Phi \text{MDE}}} - 1\). Figure (a) corresponds to the linear potential, and figure (b) is for the exponential potential. Solid lines represent numerical solutions, and dashed lines of the same color indicate corresponding fitting functions.
   }
   \label{fig:coeff/phimde}
\end{figure}

\begin{figure}
    \centering
    \begin{subfigure}[b]{0.46\textwidth}
        \centering
        \includegraphics[width=\textwidth]{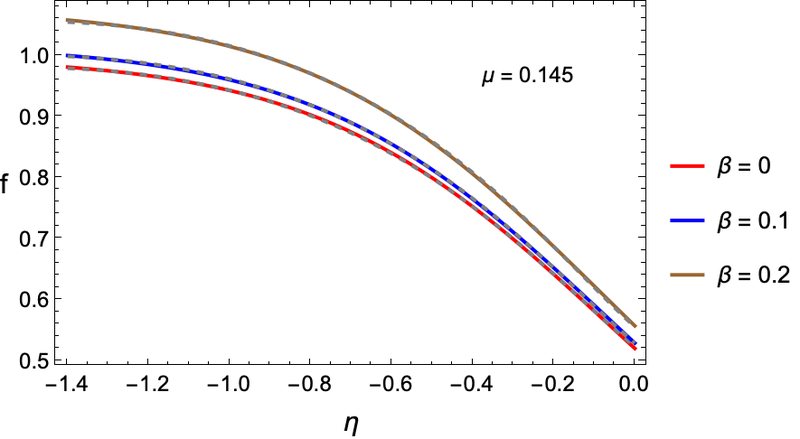}
        \caption{}
    \end{subfigure}
    \centering
    \begin{subfigure}[b]{0.46\textwidth}
        \centering
        \includegraphics[width=\textwidth]{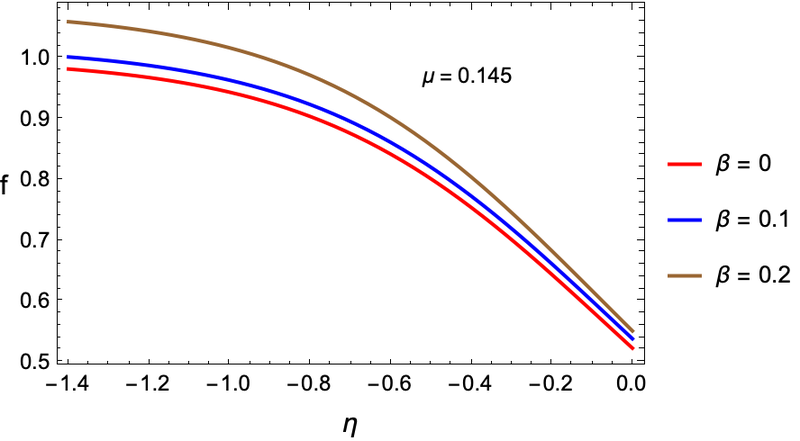}
        \caption{}
    \end{subfigure}
   \caption{Comparison between the numerical solution (solid line) and the fit function (dashed line) for the growth rate $f $. 
    Panel (a) corresponds to the linear potential, and panel (b) to the exponential potential. The initial condition of numeric solution of $f$  at $\eta=-3$  is set to  $\phi$MDE value,  which is given by $f = 1 + 2\beta^2 $.}
   \label{fig:growth_rate}
\end{figure}

\bibliography{apssamp}
\end{document}